\documentclass[preprint,12pt,longtitle,authoryear]{elsarticle}

\usepackage[margin=1.0in]{geometry} 
\usepackage{multicol}

\usepackage{epsfig}
\usepackage{latexsym}
\usepackage{epic}
\usepackage[leqno]{amsmath}
\usepackage{amssymb}

\newtheorem{theorem}{Theorem}

\newcommand{\PP}{{\mathbb P}}

\newcommand{\cM}{\mathcal M}
\newcommand{\cT}{{\mathcal T}}

\begin{document}
\begin{frontmatter}

\title{Root location in random trees: A polarity property of all sampling consistent phylogenetic models except one}
\author{Mike Steel}
\ead{mike.steel@canterbury.ac.nz}
\address{Allan Wilson Centre for Molecular Ecology and Evolution, Department of Mathematics and Statistics, University of Canterbury, Christchurch, New Zealand}



\begin{abstract}
Neutral macroevolutionary models, such as the Yule model, give rise to a probability distribution on the set of discrete rooted binary  trees over a given leaf set.  Such models can provide a signal as to the approximate location of the root when only the unrooted phylogenetic tree is known, and this signal becomes relatively more significant as  the  number of leaves grows.  In this short note, we show that among models that treat all taxa equally, and are sampling consistent (i.e. the distribution on trees is not affected by taxa yet to be included), all such models, except one,  convey some information as to the location of the ancestral root in an unrooted tree.

\end{abstract}
\end{frontmatter}

\section{Introduction}

Random neutral models of speciation (and extinction) have been a central tool for studying macroevolution, since the pioneering work of G.U. Yule in the 1920s  \citep{yule}.   Such models typically provide a  probability distribution on rooted binary trees for which the  leaf set comprises some given subset of present-day taxa.   The `shape' of these trees  has been investigated in various phylogenetic studies (see, for example, \cite{blum}) as it reflects properties of the underlying processes of speciation and extinction. Ignoring the branch lengths and considering just the topology of the trees provides not only a more tractable analysis, it also  allows for a fortuitous  robustness:  several different processes (e.g. time- or density-dependent speciation and extinction rates) lead to the same probability distribution on discrete topologies, even though the processes are quite different when branch lengths are considered \citep{ald}.

In this short paper, we are concerned with the extent to which phylogenetic models for tree topology convey information as to where the tree is rooted.  This is motivated in part by the fact that such models are used as priors in phylogenetic analysis (\cite{jon}, \cite{vel2}) and that sequence data analysed assuming the usual  time-reversible Markov processes typically returns an unrooted tree (i.e. the location of the root is unknown). Many techniques  attempt to estimate the root of the tree  using the data and additional assumptions (e.g. a molecular clock or the inclusion of an additional taxon that is known to be an `outgroup'), or using properties of the tree
that depend on branch length (e.g. `midpoint rooting', where the tree is rooted in the middle of the longest path between any two leaves) \citep{bod}.

Here, we are interested in a much more basic question:  what (if any) information the prior distribution on the topology alone itself might carry as to the location of the root of the tree.  While it has been known that some models convey root-location information, we show here a stronger result -- all models that satisfy a natural requirement (sampling consistency) have preferred root locations for a tree, except for one very special model.  We begin by recalling some phylogenetic terminology.

\subsection{Definitions and notation}

Let  $R(n)$  denote the finite set of rooted binary phylogenetic trees on the leaf set $[n]=\{1,2, \ldots, n\}$ for $n \geq 2$;  this set has size $(2n-3)!! = 1\times 3 \times 5 \times \cdots \times (2n-3)$  (see, for example, \cite{sem}).   
 
\begin{figure}[h] \begin{center}
\resizebox{12cm}{!}{
{
\includegraphics{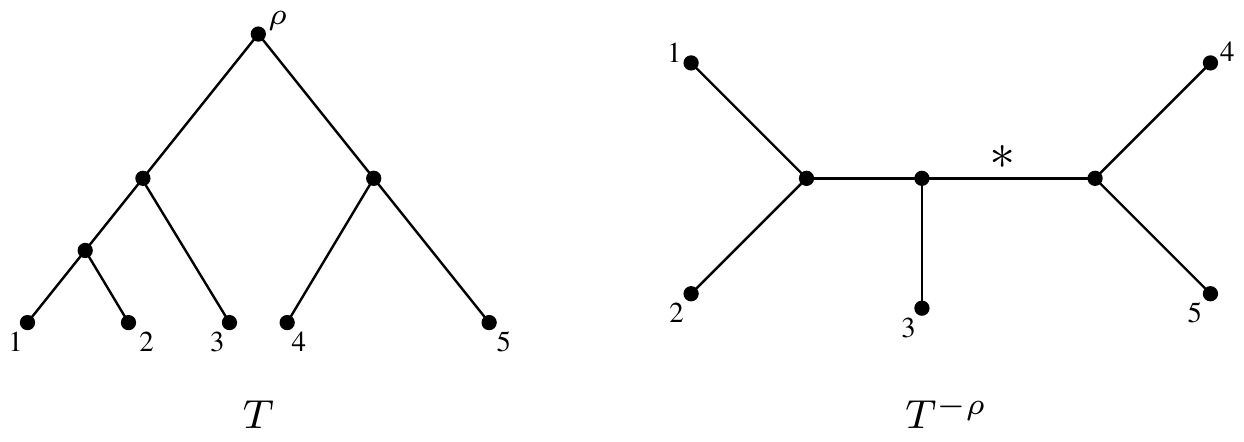}}
}
\caption{One of the 105 trees in $R(5)$ and the associated unrooted tree $T^{-\rho}$ in $B(5)$ obtained from $T$ by identifying the two edges incident with $\rho$ by a single edge (indicated by *).  Each of the seven edges of $T^{-\rho}$  corresponds to a different rooted tree.}
\end{center}
\label{fig1}
\end{figure}

 Given a tree $T \in R(n)$, let $T^{-\rho}$ denote the unrooted binary phylogenetic tree obtained by suppressing the root vertex $\rho$.  
This many-to-one association $T \mapsto T^{-\rho}$ is indicated in Fig. 1, where we have used *  to indicate the edge of $T^{-\rho}$ on which the root vertex $\rho$ would be inserted in order to recover $T$.    Notice that placing $\rho$ at the midpoint of any edge of the tree on the right in Fig.~\ref{fig1} leads to a different rooted tree.  

Thus, since an unrooted binary tree with $n$ leaves has $2n-3$ edges,  if $B(n)$ denotes the set of unrooted binary phylogenetic trees on leaf set $[n]$ then $|R(n)|= (2n-3)|B(n)|$.    

\ 

Given a subset $Y$ of $[n]$, and a (rooted or unrooted) binary  phylogenetic tree $T$ with leaf set $[n]$, let $T|Y$ denote the induced binary tree
 (root or unrooted, respectively) that connects the leaves in $Y$ (for further details, see \cite{sem}).  
 
 Now suppose we have some random process for generating a rooted binary tree on leaf set $[n]$.  We will denote the resulting randomly-generated tree as $\cT_n$.   Thus, $\cT_n$ is an element of $R(n)$ while $\cT_n^{-\rho}$ is an element of $B(n)$. To distinguish more clearly  between rooted and unrooted trees, we will often write $T_U$ to make it clear that we are referring to an unrooted tree.

\subsection{Properties of neutral phylogenetic models}

A well-studied probability distribution on $R(n)$ is the Yule--Harding model \citep{har, yule},  which can be described recursively as follows.  
Start with a tree with two (unlabelled) leaves, and repeatedly apply the following construction: For the tree constructed thus far, select a leaf uniformly at random and attach a new leaf to the edge incident with this leaf (by a new edge) and continue until the tree has $n$ leaves.  This produces a random rooted binary tree with $n$ unlabelled leaves (sometimes referred to as a `tree shape'). We then generate an element of $R(n)$ by assigning the elements of $[n]$ randomly to the leaves of this tree.

This probability distribution arises under quite general conditions, provided an exchangeability assumption is made. We describe this briefly here (for more detail, see \cite{ald}).  Consider  any model of speciation and extinction in which the rates of these two events can either be constant or vary arbitrarily with time (and even depend on the past or on the number of lineages present).  Then provided that each event (speciation or extinction) is equally likely to affect any one of the extant lineages at any given time, the resulting probability distribution on $R(n)$ is that described by the Yule--Harding model (moreover, this model also provides an equivalent distribution on tree topologies to that given by the Kingman coalescent model from population genetics, when, once again, branch lengths are ignored \citep{ald, zhu}). 

A feature of the Yule--Harding model is that given just the associated unrooted tree $T^{-\rho}$, one can readily calculate the maximum likelihood (ML) estimate of the edge(s) of $T^{-\rho}$ on which the root node $\rho$ was located (such ML edges are always incident with one of the (at most two) centroid vertices of $T^{-\rho}$); moreover,  the probability that an ML-edge contains the root node  tends to a non-zero constant ($4\log(4/3)-1 \sim 0.15$)  as $n \rightarrow \infty$ \cite{mck}.  Indeed, if we consider the edges  within three edges of this ML edge, the probability that at least one of them contains the root node is close to 0.9 \cite{mck}.  Thus, on a very large unrooted tree, $T_U$, we can isolate the likely location of the root to a relatively small proportion of edges (edges that are incident with, or near to the centroid vertex of $T_U$).  Similar results concerning the initial (root) vertex for a quite different  model of tree growth were derived in 1970 \cite{hai}.

On the other hand,  a model which provides no hint as to which edge in the associated unrooted tree might have contained the root is the {\em PDA model}  (for `proportional to distinguishable arrangements'), which is simply the  uniform probability distribution on $R(n)$. This model is not  directly described by a  model of macroevolution involving speciation and/or extinction, in the same way that the Yule--Harding distribution is, although it is possible to derive the PDA distribution
under somewhat contrived evolutionary assumptions.   These include conditioning on events such as a short window of opportunity for speciation and the survival of the tree to produce $n$ leaves \citep{mck}, conditioning on a critical binary branching process producing $n$ leaves \citep{ald}; or a general conditional independence assumptions regarding parent and daughter branches in a tree \citep{pin}.  

Notice that both  the  PDA and Yule--Harding models comprise a {\em family} of probability distributions (i.e. they provide a probability distribution on $R(n)$ for each $n \geq 2$).  Thus we will refer to any such sequence  $(p_n: n \geq 2)$ of probability distributions on $(R(n): n \geq 2$) as a  {\em phylogenetic model} $\cM$ for $(R(n): n \geq 2)$, and we will also write $\PP(\cT_n = T)$ for the probability $p_n(T)$ when $T \in R(n)$.

We now list three properties that any family of probability distributions on $(R(n): n \geq 2)$ can posses. The first two (the exchangeability property (EP) and sampling consistency (SC)) are satisfied by several models (including  the PDA and Yule--Harding model), while the third, root invariance (RI), does not hold for the Yule--Harding model, as we saw above, however it does hold for the PDA model.

\newpage
\begin{itemize}
\item[{\bf EP}] [{\em Exchangeability property}] For each $n\geq 2$, if $T$ and $T'$ are trees in $R(n)$ and $T'$ is obtained from $T$ by permuting its leaf labels, then $p_n(T') = p_n(T).$
\item[{\bf SC}]  [{\em Sampling consistency}]  For each $n \geq 2$, and  each tree $T$ in $R(n)$ we have:
$$\PP(\cT_{n+1}|[n] = T) = p_n( T).$$
\item[{\bf RI}]  [{\em Root invariance}]  For each $n\geq 2$, if $T$ and $T'$ are trees in $R(n)$ and they are equivalent up to the placement of their roots (i.e. $T^{-\rho} = T'^{-\rho}$)  then $p_n(T') = p_n(T).$\end{itemize}
\bigskip

The exchangeability property (EP), from \cite{ald},  requires the probability of a particular phylogenetic tree to depend just on its  shape and not on how its leaves are labelled (this property is called `label-invariance' in \cite{penny}).

The  sampling consistency (SC) property, also from \cite{ald},  states that the probability distribution on trees for a given set of taxa does not change if we add another taxon (namely, $n+1$) and consider the induced marginal distribution on the original set of taxa.  The condition seems reasonable if we wish the model to be `stable' in that sense that, in the absence of any data associated with the taxa,  $p_n$ should depend just on the taxa present, and not on taxa yet to be discovered (or not included in the set of taxa under study)\footnote{Note that in SC, the probability $\PP(\cT_{n+1}|[n] = T)$ is simply the sum of $p_n(T')$ over all $T' \in R(n+1)$ for which $T'|[n] = T$,  so SC is a linear constraint  that applies between the $p_n$ and $p_{n+1}$ values.}.

The  root-invariance (RI) property  states that the model does not prefer any particular rooting of a tree (i.e. any re-rooting of the tree would have equal probability).  

Note that the three properties EP, SC and RI pertain to distributions on trees in which the taxa have yet to have any biological data associated with them (i.e. they are `prior' to considering any particular data).   If the taxa come with data that is used to construct a probability distribution on trees, then clearly  EP will often not hold,  and SC could also fail, since data provided by an additional taxon can influence the relative support for trees on an existing set of taxa. RI may or may not fail, depending on the assumptions of the model (e.g. whether or not it is time-reversible, or whether or not a molecular clock is imposed). 

Several families of probability distributions on  $(R(n): n \geq 2)$  satisfy both EP and SC,  including the   `{\em $\beta$--splitting model}', a one-parameter family that was described by  \cite{ald} and which includes, as special cases, the Yule--Harding model (when $\beta =0$) and the PDA model (when $\beta = -3/2$).   Other phylogenetic models satisfying SC have also been studied recently in \cite{jon}.    Models satisfying EP and SC have been studied (and characterised) recently by \cite{haa} and \cite{mcc}.

If we consider the combination of EP and RI   several possible distributions on $R(n)$ satisfy these two properties, since one may simply define {\em any}  probability distribution on unrooted binary tree shapes, and extend this to labelled and rooted trees by imposing RI and EP.

Finally, consider the combination of SC and RI.  The PDA distribution satisfies these two properties, and, as noted already, it also satisfies EP.  
The point of this short note is to show that, unlike the other two combinations of properties,  apart from the PDA model, there is no other phylogenetic model that satisfies the combination SC and RI. This `impossibility' result is of similar spirit to (but is quite unrelated to) the result of   \cite{vel} concerning phylogenetic models that are uniform on clades of all given sizes.

\section{Results}

We now state the main result of this short note, the proof of which is given in the Appendix. 
\begin{theorem}
A phylogenetic model  $\cM=(p_n: n \geq 2)$ for  $(R(n): n \geq 2)$ satisfies the two properties SC and RI  if and only if $\cM$ is the PDA model.
\label{thm1}
\end{theorem}

The relevance of this theorem is that the PDA model does  not describe the shape of most published phylogenetic trees derived from biological data very well, as the latter trees are typically more balanced than the PDA model predicts  \citep{ald, ald2, blum}. Moreover, as noted already, the PDA model does not have a compelling biological motivation.  Thus, the significance of Theorem~\ref{thm1} is that any `biologically realistic'  sampling consistent distribution on discrete rooted phylogenetic trees necessarily favours some root locations over others in the associated unrooted tree topology.  And this holds without knowledge of the branch lengths or, indeed, of any data.  

As a corollary of Theorem 1,  the only value of $\beta$ for which the $\beta$--splitting model satisfies sampling consistency and root invariance is
$\beta = -3/2$, which corresponds to the PDA model. For other values of $\beta$, it may be of interest to determine how accurately one can estimate the exact (or approximate) location of the edge of an unrooted tree that contained the root node, when the rooted tree evolved under the $\beta$--splitting model. 

Finally, we note that the Theorem~\ref{thm1} does not require EP to hold at any point, however it falls out as a second consequence of the theorem that any phylogenetic model that satisfies SC and RI must also satisfy EP (since this holds for the PDA model).  The conditions EP and SC also apply to phylogenetic models for {\em un}rooted trees -- simply replace $R(n)$ with $B(n)$ in their definition -- and the reader may wonder whether Theorem~\ref{thm1} is merely a consequence of a result that states that any phylogenetic model on unrooted trees that satisfies SC is uniform. Such a result, if true, would indeed imply Theorem~\ref{thm1}, but such a result does not hold, even if we append condition EP to SC;  a simple counterexample is  the probability distribution on unrooted trees induced by the Yule--Harding model.

\subsection{Acknowledgments}
I thank Arne Mooers and Elliott Sober for some helpful comments on an earlier version of this manuscript.

\newpage
\bibliographystyle{elsarticle-harv}
\bibliography{Polarity_steel_MPE}

\section{Appendix: Proof of Theorem~\ref{thm1}.}

Since the PDA model clearly satisfies RI and SC, it remains to establish the `only if' claim. To this end, we first establish the following:
For any phylogenetic model $\cM=(p_n: n \geq 2)$ for  $(R(n): n \geq 2)$ that satisfies SC, and any $T_U \in B(n)$, the following identity holds:
\begin{equation}
\label{helps}
\PP(\cT_{n+1}^{-\rho}|[n]= T_U) = \PP(\cT_n^{-\rho} = T_U).
\end{equation}
To establish (\ref{helps}), first observe that for any tree $T_{n+1}$ in $R(n+1)$, one has $T_{n+1}^{-\rho}|[n] = (T_{n+1}|[n])^{-\rho}$.  Thus,
$$\PP(\cT_{n+1}^{-\rho}|[n]= T_U) = \PP((\cT_{n+1}|[n])^{-\rho}= T_U) = \sum_{T \in R(n): T^{-\rho} = T_U}\PP(\cT_{n+1}|[n]= T),$$
and, by SC, we can express this as: $$ \sum_{T \in R(n): T^{-\rho} = T_U}\PP(\cT_n = T) =  \PP(\cT_n^{-\rho} = T_U),$$
which establishes the claimed identity (\ref{helps}). 

Returning to the proof of Theorem \ref{thm1}, observe that, by RI, we can describe the probability distribution $p_n$ as that in equivalent to the one in which we first generate an unrooted phylogenetic tree $T_U \in B(n)$ with some associated probability $q(T_U)$ and then select one of the edges of $T_U$ uniformly at random to subdivide as the root vertex.  We will refer to this second (uniform) process as the {\em root-edge selection} process.   Thus, for any $T_U \in B(n)$,  we have:
\begin{equation}
\label{eq1}
q(T_U) = \PP(\cT_n^{-\rho} = T_U),
\end{equation}
and, for any $T \in R(n)$, we have, from RI that:
\begin{equation}           
\label{eq2}
 \PP(\cT_n = T) = \frac{1}{(2n-3)} q(T^{-\rho}).
 \end{equation}

Let $b(n) = |B(n)|$ (i.e.  $b(n)=(2n-5)!!=|R(n-1)|$). 
 We will show by induction on $n$ that $q$ is uniform on $B(n)$ for all $n \geq 2$ (i.e. $q(T_U) = \frac{1}{b(n)}$ for all $T_U \in B(n)$ and all $n \geq 2$).
This induction hypothesis holds for $n=2$,  so supposing that it holds for $n \geq 2$, we will use this to show that $q$ is uniform on $B(n+1)$.  First observe that, for $T_U \in B(n)$, we can apply identity (\ref{helps}), since SC holds, to deduce that:
\begin{equation}
\label{eq3}
\PP(\cT_{n+1}^{-\rho}|[n]= T_U) = q(T_U) = \frac{1}{b(n)},
\end{equation}
where the second equality holds by the induction hypothesis. 

Now, for each edge $e$ of $T_U$ consider the tree $T_U^e \in R(n+1)$ obtained from $T_U$ by attaching leaf $n+1$ to the midpoint of edge $e$ with a new edge.  
Let $p(e) = \PP(A_e|B_{T_U})$ where $A_e, B_{T_U}$  denote the nested events defined by: 
$$A_e:= `\cT_{n+1}^{-\rho} = T_{U}^{e}\mbox{'}   \mbox{ and } B_{T_U}:=` (\cT_{n+1}^{-\rho})|[n]= T_U\mbox{'}.$$
Thus $p$ is a probability distribution on the edges of $T_U$ and our aim is to show that $p$ is uniform. 
To this end, recall that on $T_U^e$ each edge has a uniform root-edge selection probability by RI.  This implies (by  SC)  that the 
 root-edge selection process on $T_U$ will select $e$ with the following probability:
 \begin{equation}
 \label{eq4}
p(e)\frac{3}{2n-1}+(1-p(e))\frac{1}{2n-1}.
\end{equation}
In this expression, the term $2n-1$ in the denominator is simply the number of edges of $T_U^e$, the numerator term $3$ corresponds to the three edges of $T_U^e$ consisting of the edge incident with $n+1$ and the two edges incident with that edge.  Since the root-edge selection process on the edges of $T_U$ is also uniform, and this tree has $2n-3$ edges, it follows from 
the expression in (\ref{eq4}) that:
$$p(e)\frac{3}{2n-1}+(1-p(e))\frac{1}{2n-1} = \frac{1}{2n-3}.$$
Thus, 
$p(e) = \frac{1}{(2n-3)}$, and so  $p$ is the uniform distribution on the edges of $T_U$, as claimed.

Finally, observe that the uniformity of $p$ now entails that $q$ is uniform on $B(n+1)$ since for any tree $T'_U \in B(n+1)$, we can 
select $T'_U \in B(n)$ and an edge $e$ of $T_U$ for which $T'_U=T_U^e$ and then:
$$q(T'_U) =q(T_U^e) =  \PP(\cT_{n+1}^{-\rho}= T_U^e) = \PP(A_e) =  \PP(A_e \& B_{T_U})= \PP(A_e|B_{T_U})\PP(B_{T_U}),$$
and thus: $$q(T'_U) =  \frac{1}{2n-3}\cdot\frac{1}{b(n)} = \frac{1}{b(n+1)}.$$

Thus, we have established the induction step required to show that $q$ is uniform on $B(n)$ for all $n \geq 2$.
It now follows from Eqn. (\ref{eq2})  that $p_n$ is uniform on $R(n)$, for all $n \geq 2$, which completes the proof of Theorem 1. 
\hfill$\Box$

\end{document}